\RequirePackage{lineno} 
\documentclass[aps,prc,superscriptaddress,nofootinbib,showpacs,floatfix,singlecolumn]{revtex4}
\usepackage{graphicx} 
\usepackage{float} 
\usepackage{amssymb}
\usepackage{url}
\usepackage[colorlinks,linkcolor=blue,citecolor=blue,filecolor=black,urlcolor=blue]{hyperref}
\usepackage{cancel}
\usepackage{MnSymbol}
\usepackage{multirow}
\usepackage{bm}

\newcommand{\sqrtsnn}{\mbox{$\sqrt{s_{\mathrm{NN}}}$}}

\newcommand{\pT} {p_{\mathrm{T}}}
\newcommand{\pTa} {p_{\mathrm{T1}}}
\newcommand{\pTb} {p_{\mathrm{T2}}}

\newcommand{\lr}[1]{\left\langle #1\right\rangle}

\newcommand{\Dphi}{\mbox{$\Delta \phi$}}
\newcommand{\Deta}{\mbox{$\Delta \eta$}}
\usepackage{color}
\usepackage{harpoon}

\graphicspath{{./}{./data/figs/}}
\begin{document}
\title{Robustness of principal component analysis on harmonic flow in heavy ion collisions}
 \newcommand{\sunysb}{Department of Chemistry, Stony Brook University, Stony Brook, NY 11794, USA}
 \newcommand{\bnl}{Physics Department, Brookhaven National Laboratory, Upton, NY 11796, USA}
 \newcommand{\chep}{Center for High Energy Physics, Peking University, Beijing 100871, China}
 \newcommand{\buphy}{Department of Physics and State Key Laboratory of Nuclear Physics and Technology, Peking University, Beijing 100871,China}
 \newcommand{\cicq}{Collaborative Innovation Center of Quantum Matter, Beijing 100871, China}
\author{Ziming Liu} \affiliation{\buphy}
\author{Arabinda Behera}\affiliation{\sunysb}
\author{Huichao Song}\affiliation{\buphy}\affiliation{\chep}\affiliation{\cicq}
\email{Huichaosong@pku.edu.cn}
\author{Jiangyong Jia}\affiliation{\sunysb}\affiliation{\bnl}
\email{jiangyong.jia@stonybrook.edu}
\date{\today}
\begin{abstract}
The principal component analysis (PCA), a mathematical tool commonly used in statistics, has recently been employed to interpret the $\pT$-dependent fluctuations of harmonic flow $v_n$ in terms of leading and subleading flow modes in heavy ion collisions. Using simulated data from {\tt AMPT} and {\tt HIJING} models, we show that the PCA modes are not fixed, but depend on the choice of the particle weight and the $\pT$ range. Furthermore, the shape of the leading mode is affected by the presence of non-flow correlations, and fake subleading mode may arise from the mixing of non-flow correlations with leading flow mode with a magnitude that could be larger than the genuine subleading flow mode. Therefore, the meaning of PCA modes and their relations to physical leading and subleading flow modes associated initial state eccentricities need to be further clarified/validated in realistic model simulations.
\end{abstract}
  \maketitle
\section{Introduction}\label{sec:1}
Heavy-ion collisions at RHIC and the LHC create the quark gluon plasma (QGP) whose space-time evolution is well described by relativistic viscous hydrodynamics~\cite{Heinz:2013th,Gale:2013da,Luzum:2013yya,Shuryak:2014zxa,Jia:2014jca,Song:2017wtw}. During its expansion, the large pressure gradients of the QGP convert the spatial anisotropies in the initial-state geometry into momentum anisotropies of the final-state particles, the strength of which is also sensitive to transport properties of the QGP such as the shear viscosity to entropy density ratio~\cite{Luzum:2008cw,Song:2010mg,Song:2012ua,Niemi:2015qia,Bernhard:2016tnd}.

The momentum anisotropy is described by a Fourier expansion of particle probability density in azimuthal angle $\phi$ and transverse momentum $\pT$
\begin{align}\label{eq:1}
\frac{dN}{d{\bm p}} \propto 1+2 v_n(\pT) \cos n (\phi-\Phi_n(\pT))\;,
\end{align}
where $d{\bm p}=d\pT d\phi$ is the phase space, and the $v_n$ and $\Phi_n$ represent the amplitude and phase of the $n^{\mathrm{th}}$-order flow vector $V_n=v_n(\pT) e^{{\textrm i}n\Phi_n(\pT)}$. The $V_n$ reflects the hydrodynamic response of the produced medium to the $n^{\textrm{th}}$-order initial-state eccentricity vector~\cite{Gardim:2011xv,Niemi:2012aj}, denoted by ${\mathcal{E}}_n=\epsilon_n {\mathrm e}^{{\textrm i}n\Phi^{\varepsilon}_n}$. Model calculations show that an approximate linear relation $V_n\propto{\mathcal{E}}_n$ is valid for $n=2$ (elliptic flow) and 3 (triangular flow). 

Experimentally, the $v_n$ coefficients are often extracted using two-particle correlation in relative azimuthal angle $\Dphi=\phi_1-\phi_2$, averaged over many events in a given centrality class~\cite{ATLAS:2012at}. The long-range part of the two-particle correlation function can be expressed as
\begin{align}\label{eq:2}
&\lr{\frac{dN_{\mathrm{pairs}}}{d{\bm p}_1d{\bm p}_2}} \propto 1+2\sum_{n=1}^{\infty} V_{n\Delta}(\pTa,\pTb) \cos (n \Dphi),\\\label{eq:2b}
&V_{n\Delta}(\pTa,\pTb)\equiv \lr{V_{n}(\pTa)V_{n}(\pTb)^*}=\lr{v_{n}(\pTa)v_{n}(\pTb) \cos n(\Phi_n(\pTa)-\Phi_n(\pTb))}\;.
\end{align}
The $V_{n\Delta}(\pTa,\pTb)$ is a real-valued and symmetric covariance matrix. In many traditional flow measurements, the single particle flow coefficients are defined as $v_n(\pT) \equiv \sqrt{V_{n\Delta}(\pT,\pT)}$, by assuming a factorization relation~\cite{Gardim:2012im}: $V_{n\Delta}(\pTa,\pTb)\approx v_n(\pTa)v_n(\pTb)$. This method works well if the influence from flow fluctuations is small. In reality, significant factorization breaking has been observed in Pb+Pb collisions at the LHC, especially for $v_2$ in central collisions where the effect can be as large as 20\%~\cite{Khachatryan:2015oea,Aaboud:2017tql}. Recently, it was proposed that this factorization breaking is associated with the radial excitation of eccentricity~\cite{Mazeliauskas:2015vea}, ${\mathcal{E}}^{(\alpha)}_n$ ($\alpha= 1, 2,...$) known as leading and subleading eccentricities, which characterize the modulation of the energy density along the radial direction with different length scales. After the hydrodynamic evolution, these radial excitations give rise to independent $n^{\mathrm{th}}$-order flow modes, $V_n^{(\alpha)}(\pT)$ with $\alpha=  1, 2,...$, known as leading, subleading and subsubleading flow. These modes fluctuate event-by-event with a phase uncorrelated with each other; these modes also have different $\pT$ dependent shapes~\cite{Mazeliauskas:2015efa}, which directly lead to the factorization breaking of flow harmonics. Besides, non-flow correlations from jets and dijets, unrelated to the initial state geometry, also contribute to the factorization breaking~\cite{Kikola:2011tu}.

Let's consider a more general situation where the harmonic flow in each event has many independent sources that may or may not correspond to the leading and subleading eccentricities,
\begin{align}\label{eq:4}
V_n(\pT) = \xi^1 F_n^{(1)}(\pT)+ \xi^2 F_n^{(2)}(\pT)+..
\end{align}
where the real vectors $F_n^{(\alpha)}(\pT)$ are bases that do not fluctuate from event to event, and the complex coefficients $\xi^\alpha$ capture the event-by-event fluctuations of the orientation and amplitude of the respective modes. Since these flow modes are independent, the phases of $\xi^\alpha$ are uncorrelated and the two-particle flow coefficients can be expressed as
\begin{align}\label{eq:6}
V_{n\Delta}(\pTa,\pTb) = \sum_{\alpha} \lr{\xi^\alpha\xi^{\alpha*}} F_{n}^{(\alpha)}(\pTa)F_{n}^{(\alpha)}(\pTb)\;.
\end{align}

Although it is straightforward to predict the $V_{n\Delta}$ if the sources of flow are known via Eq.~\ref{eq:6}, the reverse is not true in general. For example, if two modes have the same shape in $\pT$ but are uncorrelated event by event, i.e. $ \xi^a F_n(\pT)$ and $\xi^b F_n(\pT)$, their contributions to $V_{n\Delta}(\pTa,\pTb)$ are degenerate and can not be distinguished based on two-particle correlations alone. Another example is that the two-particle covariance matrix $V_{n\Delta}$ as defined in Eq.~\ref{eq:2b} can not distinguish between the $\pT$-dependent fluctuation from $v_n(\pT)$ or $\Phi_n(\pT)$, as long as both give rise to the same covariance matrix. However, if the physical flow modes are orthogonal to each other in $\pT$ phase space, they could be found using the statistical tool known as principal component analysis (PCA)~\cite{Mazeliauskas:2015vea},
\begin{eqnarray}\label{eq:7}
&V_{n\Delta}(\pTa,\pTb)=\sum_{\alpha}v_n^{(\alpha)}(\pTa)v_n^{(\alpha)}(\pTb)\;,\\\label{eq:8}
&\int d\pT w^2(\pT) v_n^{(\alpha)}(\pT)v_n^{(\beta)}(\pT)=\lambda_\alpha\delta_{\alpha\beta}
\end{eqnarray}
where $v_n^{(\alpha)}(\pT)$ are the eigenvectors of the two-particle covariance matrix and $w(\pT)$ is the weight for the particle. The PCA method was first applied to study the multiplicity fluctuations and anisotropic flow fluctuations~\cite{Bhalerao:2014mua}. Teaney and Mazeliauskas~\cite{Mazeliauskas:2015vea,Mazeliauskas:2015efa} studied the correspondence between leading, subleading eccentricities and the resulting flow modes using a hydrodynamics model simulation with simplified initial geometry. They found that the linear relation $V_n^{(\alpha)}\propto {\mathcal{E}}_{n}^{(\alpha)}$ work well for $\alpha = 0,1,2$, and the extracted flow modes are indeed approximately orthogonal to each other in the $\pT$ phase space. An experimental analysis by the CMS collaboration~\cite{Sirunyan:2017gyb} also extracted non-zero subleading flow in qualitative agreement with the model predictions. 

There are certain assumptions and limitations built into the PCA formalism Eqs.~\ref{eq:7} and \ref{eq:8}: 1) The extracted flow modes depend on the choice of weight factor $w(\pT)$, which is chosen as either unity or particle multiplicity in the literature. Although in principle it should be determined via realistic event-by-event hydrodynamic model simulations. 2) The extracted flow modes depend on the $\pT$ phase space over which the orthogonality is defined. The modes that are orthogonal in one $\pT$ range in general are not orthogonal when the range is enlarged or reduced. 3) the PCA method assumes the covariance matrix $V_{n\Delta}(\pTa,\pTb)$ are positive definite, i.e. $\lambda_\alpha>0$ in Eq.~\ref{eq:8}. This may not be the case in the presence of non-flow or long-range correlations that are not single-particle origin. Even if these contributions are not enough to change the sign of $\lambda_\alpha$, they could still affect the extracted $v_n^{(\alpha)}(\pT)$ in a non-trivial way. 

The main goal of this paper is to study the impact of these three effects on PCA method, and discuss possible limitations on the interpretation of the extracted modes. The study is carried out using simulated data that contain flow (AMPT model) and non-flow ({\tt HIJING} model). The paper is organized as follows. Section~\ref{sec:2} discusses the setup of the model and the PCA procedure. Section~\ref{sec:3} presents the main findings from the three types of tests. Section~\ref{sec:4} gives a short summary.

\section{Test setup}\label{sec:2}
We first give a brief summary of the PCA procedure. In heavy ion collisions, the number of particles $M$ produced in each event is finite. Therefore, harmonic flow vector $V_n$ can only be estimated from the observed flow vector $Q_n$ or per-particle normalized flow vector $q_n$:
\begin{align}
\label{eq:9a}
 Q_n \equiv \sum_i  e^{in\phi_i} = Q_n e^{in\Psi_n},\; {\bm q}_n \equiv \frac{\sum_i  e^{in\phi_i}}{M}= q_n e^{in\Psi_n}
\end{align}
where the sum runs over the particles in the event, $\phi_i$ are their azimuthal angles and $\Psi_n$ is the observed event plane. The magnitude and direction of ${\bm q}_n$ differ from those for the truth flow in each event, due to azimuthal fluctuations associated with finite particle multiplicity. The two-particle flow matrix is obtained as,
\begin{align}\label{eq:9}
V_{n\Delta}^q(\pTa,\pTb)\equiv\lr {{\bm q}_n(\pTa){\bm q}_n^*(\pTb)}\;
\end{align}

Principal Component Analysis (PCA) is a statistical method, which transforms correlation matrix into a set of uncorrelated single-particle modes called principal components. This is achieved by minimizing the variance of the correlation matrix to find the principal components order by order, which gives
\begin{align}\label{eq:9b}
V_{n\Delta}^q(\pTa,\pTb)=\sum_{\alpha}v_n^{q(\alpha)}(\pTa)v_n^{q(\alpha)}(\pTb)\;,
\end{align}
For sufficient number of events, the statistical fluctuations average out and the two-particle correlation matrix are expected to converge on the true flow signal. Therefore the $v_n^{q(\alpha)}(\pT)$ should be the same as $v_n^{(\alpha)}(\pT)$ within the statistical uncertainty. 

In an alternative approach, the two particle flow matrix is obtained with $Q_n$,
\begin{align}\nonumber
V_{n\Delta}^Q(\pTa,\pTb)&\equiv\lr {Q_n(\pTa)Q_n^*(\pTb)}\\\label{eq:10}
&\approx M(\pTa)M(\pTb)V_{n\Delta}^q(\pTa,\pTb)
\end{align}
The PCA decomposition is performed on $V_{n\Delta}^Q$,
\begin{align}\label{eq:11}
V_{n\Delta}^Q(\pTa,\pTb)=\sum_{\alpha}M(\pTa)v_n^{Q(\alpha)}(\pTa)M(\pTb)v_n^{Q(\alpha)}(\pTb)
\end{align}

The main difference between the two approaches are the choice of the weight factor $w(\pT)$ for the orthogonality (Eq.~\ref{eq:8}): $w(\pT)=1$ in the first approach while it is  $w(\pT)=M(\pT)$ in the second approach. Since there are more particles at low $\pT$, the second approach places more weight in the low $\pT$ region. In contrast, the first approach assigns equal weight in all $\pT$ bins. A recent work~\cite{Gardim:2019iah,Hippert:2019swu} argues that PCA modes extracted from normalized scheme ($w(\pT)=1$) seem more accurately reflect the underlying flow modes.

To understand the performance and potential limitations of the PCA method, we carried out a simulation study using the {\tt AMPT}~\cite{Lin:2004en} and {\tt HIJING}~\cite{Wang:1991hta} models. The {\tt HIJING} model combines the fluctuating initial geometry based on Glauber model and particle productions based on lund-string dynamics at low $\pT$ and  hard QCD interaction at high $\pT$. The {\tt AMPT} model starts from the particles produced by {\tt HIJING}, breaks them into partons (``string-melting'') and runs them through partonic transport. The partonic transport processes generate significant collective flow which has been demonstrated to qualitatively describe the harmonic flow $v_n$ in $p$+A and A+A collisions~\cite{Xu:2011jm,Xu:2011fe}. The fluctuating initial geometry and partonic transport processes in the {\tt AMPT} model naturally generate leading and subleading flow modes, which can be quantified via the PCA analysis similar to the data. On the other hand, the PCA analysis can also be performed using the {\tt HIJING} events to quantify the influence of the non-flow correlations.

The {\tt HIJING} and {\tt AMPT} data used in this study are generated for Au+Au collisions with impact parameter $b<3$~fm at $\sqrt{s_{\mathrm{NN}}}=200$ GeV, which corresponds to approximately 0--4\% most central events. We choose to generate ultra-central events in {\tt AMPT}, because those events are known to have relatively large subleading flow modes. The {\tt HIJING} events, on the other hand, provide a reasonable estimate of the non-flow effects. All charged particles within $0.1<\pT<3$ GeV/$c$ and $|\eta|<3$ are used. These particles are divided into seven intervals in $\pT$: 0.1--0.3, 0.3--0.5, 0.5--0.7, 0.7--1.0, 1.0--1.5, 1.5--2.0, and 2.0-3.0 GeV/$c$. In order to reduce the non-flow contributions, the particles are divided into two subevents separated in pseudorapidity: $\eta>0.4$ for subevent a and $\eta<-0.4$ for subevent b. The flow vectors are calculated in each $\pT$ bin in both subevents, and the flow covariance matrix $V_{n\Delta}$ are obtained via the standard two-subevent method~\cite{Jia:2017hbm},
\begin{align}\nonumber
&V_{n\Delta}^Q(\pTa,\pTb) = \lr {Q_n^a(\pTa)Q_n^{b*}(\pTb)}\;,\\\label{eq:12}
&V_{n\Delta}^q(\pTa,\pTb) = \lr {{\bm q}_n^a(\pTa){\bm q}_n^{b*}(\pTb)}\;.
\end{align}
The $V_{n\Delta}^{Q/q}$ are $7\times7$ covariance matrices on which the PCA decomposition is performed according to Eqs.~\ref{eq:9b} or \ref{eq:11}, respectively. By comparing $v_n^{q(\alpha)}(\pT)$ and $v_n^{Q(\alpha)}(\pT)$, we can quantify the sensitivity on the particle weights. 

The second goal of our paper is to study the sensitivity on the $\pT$ phase space. This is achieved by dropping a subset of the $\pT$ bins to obtain a covariance matrix with reduced rank, on which we perform the PCA analysis. The results obtained in reduced $\pT$ phase space are then compared with each other to quantify the stability of the leading and subleading flow modes. 

The third goal of our paper is to understand how the bias of non-flow correlations affect the extracted flow modes. This goal is achieved by constructing a two-particle correlation matrix by mixing the matrices from {\tt AMPT} and {\tt HIJING}. 
\begin{equation}\label{eq:13}
V_{n\Delta}^{\mathrm{mix}} = f V_{n\Delta}^{A} + (1-f) V_{n\Delta}^{H}
\end{equation}
where $V_{n\Delta}^{A}$ and $V_{n\Delta}^{H}$ are covariance matrices generated from {\tt AMPT} and {\tt HIJING}, respectively. By varying the fraction $f$, we can see how the non-flow affects the leading and subleading modes.

In the following sections, we present dependence of the PCA modes on particle weights, $\pT$ phase space and non-flow contributions. 

\section{Results}\label{sec:3}
\begin{figure}[h!]
\includegraphics[width=0.8\linewidth]{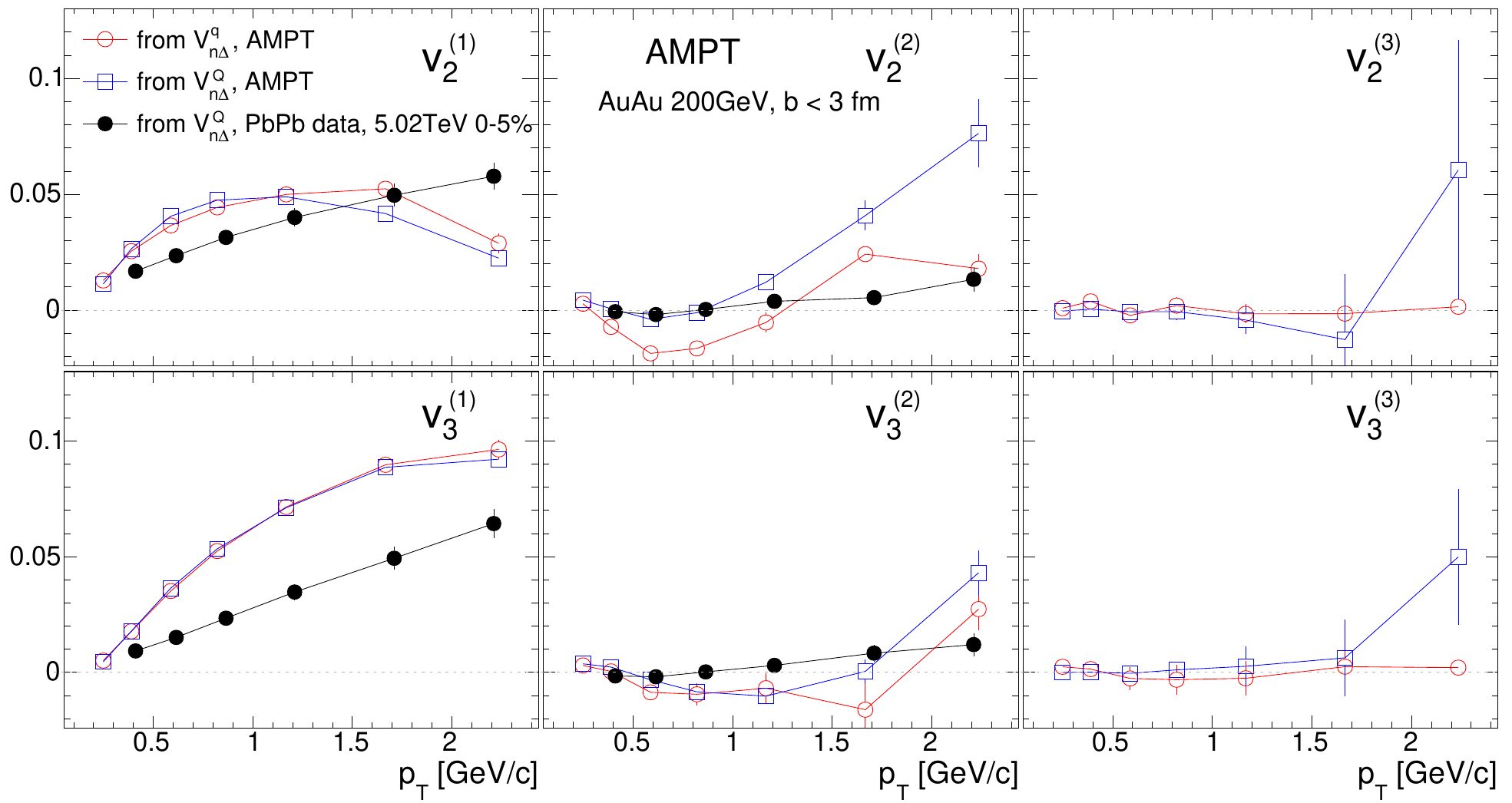}
\caption{\label{fig:1} The leading (left), subleading (middle), and subsubleading (right) modes of $v_2$ (top) and $v_3$ (bottom) obtained from PCA for central Au+Au  {\tt AMPT} events with two different normalization schemes (open circles and open squares). They are compared with results obtained for 0-5\% central Pb+Pb data from CMS Collaboration~\cite{Sirunyan:2017gyb}.}
\end{figure}

Figure~\ref{fig:1} shows the first three PCA modes for $v_2$ and $v_3$ compared between the two normalization schemes for $V_{n\Delta}$: $v_n^{q(\alpha)}$ for which different $\pT$ bin has equal weight and $v_n^{Q(\alpha)}$ for which each particle has the same weight. The subsubleading flow $v_n^{(3)}$ is largely consistent with zero within current statistical precision, and its overall magnitude is much smaller than the leading flow $v_n^{(1)}$ and subleading flow $v_n^{(2)}$. Therefore, our discussion focuses mainly on the leading and subleading flows.

Significant non-zero subleading flow $v_n^{(2)}$ are observed in Fig.~\ref{fig:1}, and in some cases, they are larger than the leading flow $v_n^{(1)}$ at high $\pT$. This indicates significant factorization breaking for the covariance matrices $V_{n\Delta}(\pTa,\pTb)$. The subleading flow $v_n^{(2)}$ always crosses zero at some intermediate $\pT$, which is the consequence of it being orthogonal with leading flow $v_n^{(1)}$. Comparing between the two normalization schemes, the crossing point is much larger for $v_n^{q(2)}$ than that for $v_n^{Q(2)}$, and the leading flow $v_n^{q(1)}$ is slightly larger than $v_n^{Q(1)}$ at high $\pT$. These differences are expected since $v_n^{q(\alpha)}$ is more affected by high $\pT$ region of the $V_{n\Delta}$ than $v_n^{Q(\alpha)}$. 

Figure~\ref{fig:1} also compares with the $v_n^{Q(\alpha)}$ results from the CMS Collaboration for 0--5\% most central Pb+Pb collisions at $\sqrtsnn=5.02$ TeV. Such comparison can be carried out because $v_n(\pT)$ for inclusive hadron is known to have very little $\sqrtsnn$ dependence, and they are consistent between Au+Au at RHIC energies and Pb+Pb at LHC energies~\cite{Aamodt:2010pa}. For the leading mode, the CMS data have different $\pT$ dependence trends from the {\tt AMPT} and the values agree within a factor of 2. For the subleading mode, the relative differences between data and {\tt AMPT} are much larger. 

Because the results depend on the weighting schemes, one has to rely on event-by-event hydrodynamic model calculations to inform us which scheme reflects better the true flow modes. A recent work~\cite{Hippert:2019swu} argue that the normalized scheme (Eq.~\ref{eq:9b}) has better correlation with the leading and subleading eccentricities. However, it is important to check if and how these schemes are sensitive to the choice of $\pT$ phase space and the non-flow effects, which are discussed as follows.

The flow modes extracted from the PCA method depend on the $\pT$ phase space on which the covariance matrix $V_{n\Delta}(\pTa,\pTb)$ is constructed. This dependence can be evaluated by systematically dropping a subset of the seven $\pT$ bins, $\pT^{1},\pT^{2},...,\pT^{7}$, and repeating the PCA decomposition. Three ``drop schemes'' labelled as A,B,C, are implemented. In scheme A, we only keep the lower $\pT$ bins, i.e. keeping $\pT^{1},...,\pT^{6-i},\pT^{7-i}$ with $i=0$--6, resulting a $(7-i)\times(7-i)$ covariance matrix $V_{n\Delta}$. In scheme B, we only keep the higher $\pT$ bins, i.e. keeping $\pT^{i+1},\pT^{i+2},...,\pT^{7}$ with $i=0$--6, resulting a $(7-i)\times(7-i)$ covariance matrix. In scheme C, we reject only one $\pT$ bin, i.e. keeping $\pT^{1},\pT^{2},...,\pT^{i},\pT^{i+2},...,\pT^{7}$ with $i=0$--6, resulting a $6\times6$ covariance matrix.  The extracted flow modes are then compared in the overlapping $\pT$ region to quantify the sensitivity to the $\pT$ bins dropped in the analysis.
\begin{figure}
\includegraphics[width=0.8\linewidth]{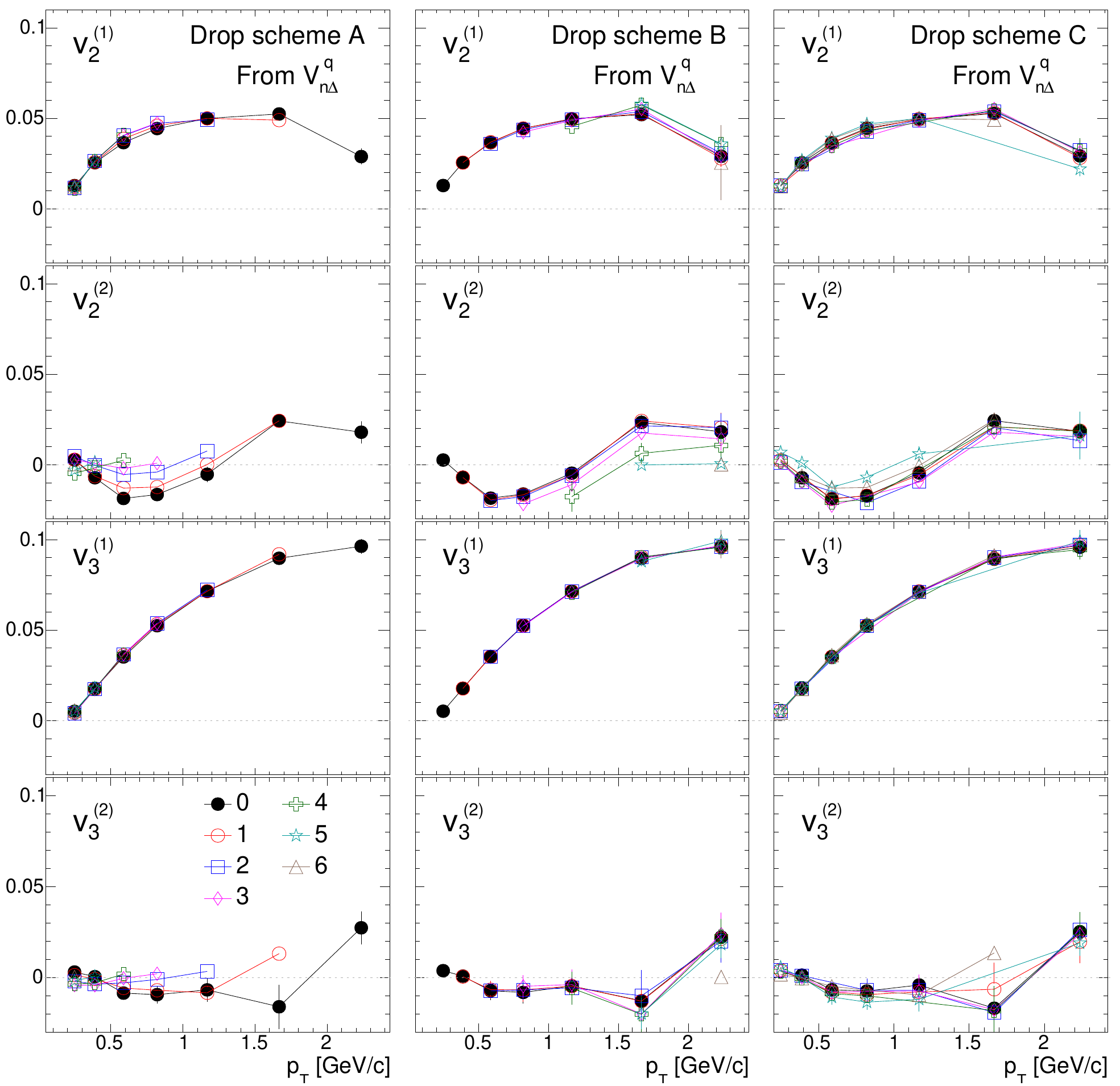}
\caption{\label{fig:2} The PCA modes from covariance matrix $V_{n\Delta}^q(\pTa,\pTb)$ Eq.~\ref{eq:12} for three $\pT$-bin dropping schemes (see text for explanation), scheme A (left column), scheme B (middle column), scheme C (right column). They are shown for $v_2^{(1)}$, $v_2^{(2)}$, $v_3^{(1)}$ and $v_3^{(2)}$ from top to bottom panels. Each panel shows 7 cases corresponding to $i=0$--6 (see text).}
\end{figure}

\begin{figure}
\includegraphics[width=0.8\linewidth]{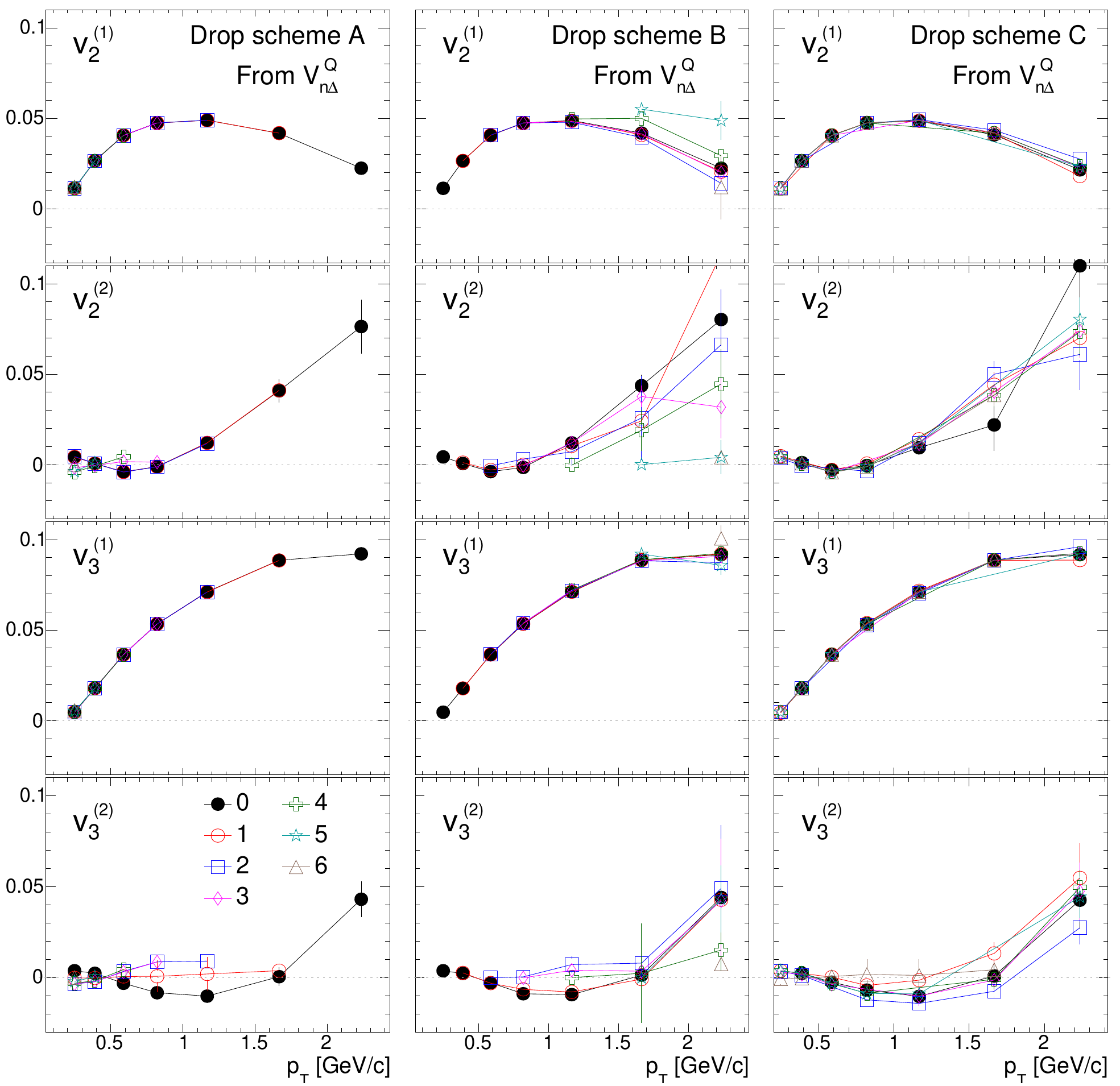}
\caption{\label{fig:3} The PCA modes from covariance matrix $V_{n\Delta}^Q(\pTa,\pTb)$ Eq.~\ref{eq:12} for three $\pT$-bin dropping schemes (see text for explanation), scheme A (left column), scheme B (middle column), scheme C (right column). They are shown for $v_2^{(1)}$, $v_2^{(2)}$, $v_3^{(1)}$ and $v_3^{(2)}$ from top to bottom panels. Each panel shows 7 cases corresponding to $i=0$--6 (see text).}
\end{figure}

The left column of Fig.~\ref{fig:2} shows the leading and subleading modes obtained from scheme A. The leading mode $v_n^{(1)}$ has weak dependence, while the subleading mode $v_n^{(2)}$ has much stronger dependence on the number of $\pT$ bins used. When only the low $\pT$ bins are used, the $v_2^{(1)}$ increases slightly, the $v_3^{(1)}$ is nearly unchanged, while the $v_2^{(2)}$ and $v_3^{(2)}$ are closer to zero. The results from scheme B are shown in the middle column of Fig.~\ref{fig:2}.  When only the high $\pT$ bins are used, the $v_2^{(2)}$ decreases significantly, while the others only change slightly. The right column of Fig.~\ref{fig:2} shows the results from scheme C where only one $\pT$ bin is dropped. The leading mode $v_n^{(1)}$ is rather insensitive to which $\pT$ bin to drop, while the subleading mode is much more sensitive. 

Figure~\ref{fig:3} show the same results obtained for $v_n^{Q(\alpha)}$. The leading modes are relatively insensitive to the dropping scheme, except for the $v_2^{(1)}$ in the scheme B. The subleading modes are much more sensitive to the phase space dropped in all three schemes.

The results from Figs.~\ref{fig:1}--\ref{fig:3} show that the PCA modes are dependent on the choice of normalization scheme for the flow vector as well as the $\pT$ phase space chosen for $V_{n\Delta}$.

\begin{figure}
 \includegraphics[width=1\linewidth]{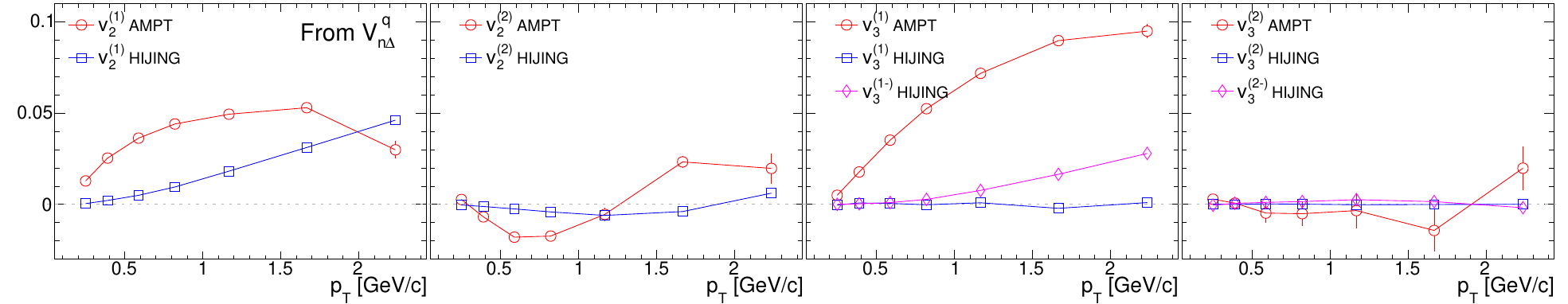}\\
 \includegraphics[width=1\linewidth]{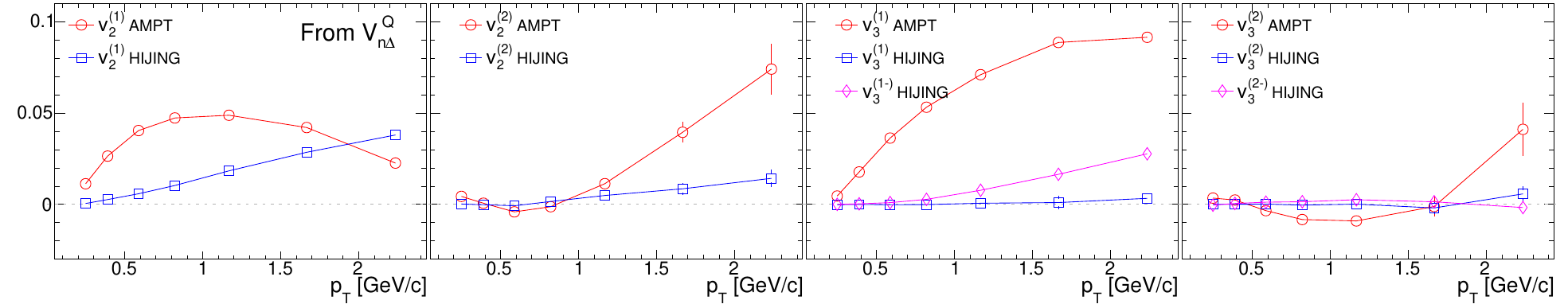}
 \caption{\label{fig:4} The PCA modes, $v_2^{(1)}$ (left column), $v_2^{(2)}$ (second column), $v_3^{(1)}$ (third column) and $v_3^{(2)}$ (right column), obtained from $V_{n\Delta}^q(\pTa,\pTb)$ (top) and $V_{n\Delta}^Q(\pTa,\pTb)$ (bottom) from {\tt HIJING} (open circles) and {\tt AMPT} (open squares). The PCA modes $v_3^{(\alpha)}$ from {\tt HIJING} are nearly zero due to negative-definiteness of $V_{3\Delta}^q$, therefore the PCA modes for $-V_{3\Delta}^q$ are also shown (open diamonds).}
\end{figure}

Next we study how non-flow correlations influence the PCA modes, where non-flow correlations are simulated using the {\tt HIJING} model. As mentioned in Section~\ref{sec:2}, we calculate the covariance matrices $V_{n\Delta}^q (\pTa,\pTb)$ obtained via two-subevent method with $\Deta$ gap of 0.8 unit. In this case, the $V_{n\Delta}^q$ from {\tt AMPT} are dominated by flow while $V_{n\Delta}^q$ from {\tt HIJING} are dominated by non-flow associated with large away-side dijet correlations. Figure~\ref{fig:4} compares the leading and subleading modes for {\tt AMPT} and {\tt HIJING}. The {\tt HIJING} model shows large leading and subleading modes for $v_2$, but nearly zero signal for $v_3$. The peculiar behavior for the $v_3$ can be understood from the nature of the non-flow correlation from dijets. Due to the $\Deta$ gap requirement, the dijets mainly produce pairs at $\Dphi\sim\pi$, whose contribution to $V_{n\Delta}^q$ is positive for even harmonics and negative for odd harmonics~\footnote{Fourier coefficients for a delta function at $\Delta\phi=\pi$ are $V_{n\Delta}=\frac{1}{2\pi}(-1)^n$.}. Therefore $V_{2\Delta}^q$ values are positive and they can be decomposed to leading and subleading flow modes. In contrast, we found that the $V_{3\Delta}^q$ values from {\tt HIJING} are mostly negative, and the eigenvalues of the corresponding covariance matrix $V_{3\Delta}^q (\pTa,\pTb)$ are mostly negative (i.e. $\lambda_\alpha<0$ in Eq.~\ref{eq:8}). In this case, the $V_{3\Delta}^q$ can not be decomposed into single-particle modes, and resulting PCA modes are all zero. Therefore, in order to compare the magnitude of non-flow with flow, we extract the PCA modes from $-V_{3\Delta}^{q}(\pTa,\pTb)$, and the results are represented by $v_3^{(\alpha-)}$ ($\alpha=1,2$) in Figure~\ref{fig:4}. Indeed, significant $v_3^{(1-)}$ values are obtained but with a magnitude smaller than the $v_3^{(1)}$ from the {\tt AMPT}. This implies that although PCA is unable to find modes with negative eigenvalues for the covariance matrix from {\tt HIJING}, the non-flow effects could still influence the extracted PCA modes when both flow and non-flow are present. Although such influence is small for triangular flow modes in central A+A collisions, it is expected to become much more significant in mid-central or peripheral A+A collisions, where the $v_3^{(1-)}$ and $v_3^{(2-)}$ are expected to be much larger. 

\begin{figure}
\includegraphics[width=1\linewidth]{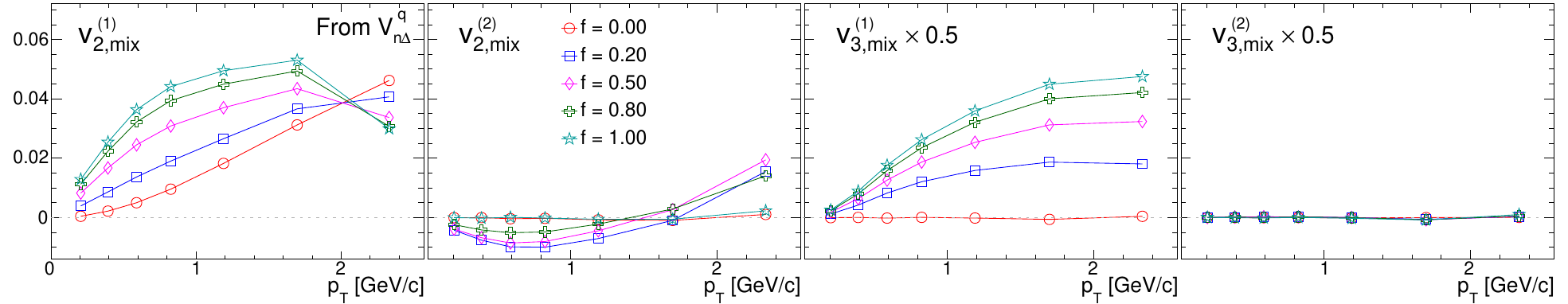}\\
\includegraphics[width=1\linewidth]{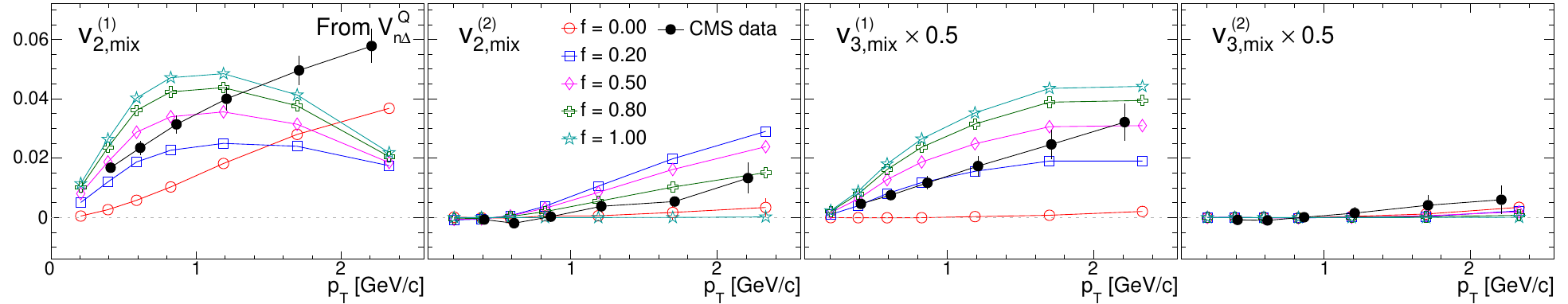}
\caption{\label{fig:5} The PCA modes, $v_{2,\mathrm{mix}}^{(1)}$ (left column), $v_{2,\mathrm{mix}}^{(2)}$ (second column), $v_{3,\mathrm{mix}}^{(1)}$ (third column) and $v_{3,\mathrm{mix}}^{(2)}$ (right column), obtained from mixed covariance matrix via Eq.~\ref{eq:13} for $V_{n\Delta}^q(\pTa,\pTb)$ (top) and $V_{n\Delta}^Q(\pTa,\pTb)$ (bottom). Different mixing fractions are shown in each panel. In the bottom panel, results are also compared with the 0-5\% central Pb+Pb data from CMS Collaboration~\cite{Sirunyan:2017gyb}.}
\end{figure}

To understand the behavior of PCA modes when both flow and non-flow are present, we proceed to mix the covariance matrices from {\tt AMPT} and {\tt HIJING} according to Eq.\ref{eq:13} but only consider the contribution of leading modes to simplify the discussion,
\small{\begin{align}\nonumber
&V_{2\Delta}^{\mathrm{mix}} = f v_{2,A}^{(1)}(\pTa)v_{2,A}^{(1)}(\pTb) + (1-f) v_{2,H}^{(1)}(\pTa)v_{2,H}^{(1)}(\pTb)\\\label{eq:14}
&V_{3\Delta}^{\mathrm{mix}} = f v_{3,A}^{(1)}(\pTa)v_{3,A}^{(1)}(\pTb) - (1-f) v_{3,H}^{(1-)}(\pTa)v_{3,H}^{(1-)}(\pTb),
\end{align}}\normalsize
where $v_{n,A}^{(1)}(\pTa)$ are leading mode of the covariance matrix from {\tt AMPT},  $v_{2,H}^{(1)}$ ($v_{3,H}^{(1-)}$) is the leading mode of $V_{2\Delta}^{H}$ (-$V_{3\Delta}^{H}$) from {\tt HIJING}, respectively. We then apply PCA method on the mixed covariance matrix $V_{n\Delta}^{\mathrm{mix}}(\pTa,\pTb)$ to obtain the leading and subleading modes.

Figure~\ref{fig:5} shows the leading and subleading modes obtained from the mixed covariance matrix. Both leading modes $v_{2,\mathrm{mix}}^{(1)}$ and $v_{3,\mathrm{mix}}^{(1)}$ smoothly interpolate between leading modes from {\tt AMPT} and leading modes from {\tt HIJING}. The shapes of leading modes as a function of $\pT$ are clearly influenced by the presence of non-flow for $v_{2,\mathrm{mix}}^{(1)}$, but are only slightly influenced for $v_{3,\mathrm{mix}}^{(1)}$. A large non-zero subleading flow mode $v_{2,\mathrm{mix}}^{(2)}$ emerges as the result of the mixing, and its magnitude shows a non-monotonic behavior as a function mixing fraction $f$. This is because the $v_{2}^{(1)}$ from the two models have different shape as a function of $\pT$. The PCA procedure naturally decomposes the mixed matrices into new orthogonal bases, leading to non-zero subleading mode. The magnitude of this ``fake'' subleading mode could be much larger than that from the CMS data as shown in the bottom middle panel. This implies that the presence of the non-flow from dijets may mix with the leading mode of elliptic flow to contaminate the subleading mode of elliptic flow. On the other hand, the mixing procedure for triangular flow does not give rise to a subleading mode, which is because the non-flow from {\tt HIJING} $v_{3,H}^{(1-)}$ is small and its shape as as function of $\pT$ is not very different from $v_{3,A}^{(1)}$ for {\tt AMPT}.

\section{Discussion and summary}\label{sec:4}
The principal component analysis (PCA), a popular mathematical tool in statistics, has been used recently to extract the leading and subleading elliptic flow $v_2$ and triangular flow $v_3$ in heavy ion collisions. In this method, one first constructs a set of two-particle azimuthal correlators $V_{n\Delta}=\lr{\cos n(\phi_1-\phi_2)}$ for particles selected in a given transverse momentum $\pT$ range, the full two-dimensional covariance matrix $V_{n\Delta}(\pTa,\pTb)$ is then decomposed into leading mode $v_n^{(1)}$ and subleading mode $v_n^{(2)}$ via the PCA procedure. We carried out this study using central Au+Au collisions generated with {\tt AMPT} and {\tt HIJING} models. The {\tt AMPT} model provides a realistic estimation of the flow signal, while {\tt HIJING} model provides a reasonable estimation of the non-flow associated with away-side dijets.

The leading and subleading flow modes $v_n^{(2)}$ ($n=2$,3) have been measured by CMS collaboration, which have been interpreted as hydrodynamic response to different radial excitations of the eccentricities in the initial state. However, it is unclear how well the PCA modes represent the true flow modes. The goal of our paper is to show that the PCA method has several limitations and the extracted modes may not precisely correspond to the true flow modes. Three types of studies are performed to check the stability of the PCA method, 1) the $V_{n\Delta}$ data in each $\pTa$ and $\pTb$ bin are chosen to have the unit weight or particle multiplicity as the weight, 2) the $V_{n\Delta}$ data are constrained to different phase space in $\pTa$ and $\pTb$, 3) the $V_{n\Delta}$ from {\tt HIJING} and {\tt AMPT} models are mixed by different fractions to vary the amount of non-flow correlations. 

While leading modes are found to be relatively stable against variation for the particle weight and the $\pT$ phase space, the subleading modes are found to be much more sensitive to these variations. This already implies that as a mathematical tool, PCA do not exactly determine the subleading flow associated with subleading eccentricities.
The leadings modes from {\tt AMPT} and {\tt HIJING}, $v_{n,A}^{(1)}$ and $v_{n,H}^{(1)}$, are found to have different shapes as a function of $\pT$. These leading modes are mixed together to form a combined $V_{n\Delta}$, on which the PCA decomposition is applied. The shape of the leading mode from the combined $V_{n\Delta}$, $v_{n,\mathrm{mix}}^{(1)}$, is sensitive to the non-flow correlations. What is more interesting is that a non-zero subleading mode, $v_{n,\mathrm{mix}}^{(2)}$, emerges from the combined $V_{n\Delta}$. The magnitude of the subleading elliptic flow $v_{2,\mathrm{mix}}^{(2)}$ can easily exceed the values from the experimental data. This implies that the presence of non-flow could mix with the leading elliptic flow mode to mimick subleading elliptic flow mode. Therefore one should be a bit cautious in the interpretation of the results given by the PCA procedure. 

Our main point then is that as a mathematical tool, PCA can not determine exactly the subleading flow associated with subleading eccentricities. The true orthogonality condition will depend on the choice of weight and phase space, as well as the initial collision geometry and the transport properties of the specific hydro model implementation. Further realistic event-by-event hydrodynamic model simulations are required to establish the relationship between leading and subleading PCA modes and leading and subleading eccentricities in the initial state.

We would like to thank the fruitful discussions with A.~Mazeliauskas, D.~Teaney and M.~Zhou. J.J and A.B acknowledge support by NSF under grant number PHY-1613294 and PHY-1913138, Z.L. and H.S. acknowledge the support by the NSFC and the MOST under grant Nos. 11675004, 11435001 and 2015CB856900 and the computing resources provided by the Super-computing Center of Chinese Academy of Science (SCCAS), Tianhe1A from the National Supercomputing Center in Tianjin, China and the High-performance Computing Platform of Peking University.
\bibliography{PCA}
\bibliographystyle{apsrev4-1}
\end{document}